\documentclass{elsart_2}

\usepackage{graphicx} 
 
\usepackage{amssymb} 
 
\begin{document} 
 
\begin{frontmatter} 
 
 
 
\title{Substructures in lens galaxies: PG1115+080 and B1555+375, two fold configurations} 
 
 
\author {Marco Miranda and Philippe Jetzer}      
 
\address{Institute for Theoretical Physics, University of Z\"urich,       
Winterthurerstrasse 190, CH-8057      
  Z$\ddot u$rich, Switzerland} 
 
\begin{abstract} 
 We study the anomalous flux ratio which is observed in some  
four-image lens systems, where the source lies close to a fold caustic.   
In this case two of the images are close to the critical curve and    
their flux ratio should be    
equal to unity, instead in several cases   
the observed value differs significantly.    
The most plausible solution is to invoke the presence of substructures,  
as for instance predicted by the Cold Dark Matter scenario,     
located near the two images.   
In particular, we analyze the two fold lens systems PG1115+080 and B1555+375,       
for which there are not yet satisfactory models which explain the observed   
anomalous flux ratios.      
We add to a smooth lens model, which reproduces well   
the positions of the images but not the anomalous fluxes,    
one or two substructures described as   
singular isothermal spheres.   
For PG1115+080 we consider a smooth model with the influence of the
group of galaxies described by a SIS and a substructure with mass $\sim
10^{5}\ M_{\odot}$ as well as a smooth model with an external shear and one substructure with mass $\sim
10^{8}\ M_{\odot}$. For B1555+375 either a strong external shear or two 
substructures with mass $\sim 10^{7}\ M_{\odot}$ reproduce the data quite well. 
       
\end{abstract} 
 
\begin{keyword} 
 cosmology: theory -- dark matter -- gravitational lensing -- galaxies:    
haloes -- substructures     
 
\end{keyword} 
 
\end{frontmatter} 
 
\section {Introduction}      
The standard lens models, although reproduce in general the relative positions    
of the images quite accurately, often have difficulties explaining the     
relative fluxes of multiply-imaged sources.     
Several possible explanations have been considered in the literature, the most    
plausible being that the lensing potential of real galaxies are not    
fully described by the simple lens models used to compute the lens    
characteristics.      
The most often invoked solution is to consider additional small-scale    
structures, which if located near the images can modify    
significantly the observed flux ratio between different images,    
in particular the so-called cusp or fold relations.    
    
The presence of substructures is naturally expected within     
the Cold Dark Matter (CDM) model, which    
has been successful in explaining a large variety of observational results     
like the large scale structure of galaxies on scales larger than 1 Mpc or     
the fluctuations of the CMB ({  Spergel et al 2003}).      
However, one of the predictions of this scenario      
is a distribution of matter, with a large number of      
small-mass compact dark matter (DM subhalos) halos, both within virialized      
regions of larger halos (Moore et al. 1999, Klypin et al. 1999)        
and in the field (DM extragalactic halos, Metcalf 2005).  At the      
same time, the observed number of dwarf galaxy satellites in the Local      
Group is more than an order of magnitude smaller than expected.        
Many theoretical studies suggest models to reduce the      
abundance of substructure      
or to suppress star formation in small clumps via astrophysical      
mechanisms (as feedback), making them dark (Bullock, Kravtosov
\& Weinberg 2000, {  Kravtsov et al. 2004, Moore et al. 2006}).      
Anyway, if the CDM paradigm is correct, we expect $\sim 10-15\%$     
of the mass of      
a present-day galactic halo ($\sim 10^{12} M_{\odot} $) within the virial      
radius to be in substructures with mass $\geq 10^7 M_{\odot}$.       
Thus in the CDM model anomalous flux ratios should be common.       
      
At present, the only way to detect these subclumps       
is through gravitational lensing, which is directly sensitive to the mass.      
This is because substructures (like globular clusters, gas clouds or       
satellite galaxies) can strongly modify the fluxes of lensed images      
relative to those predicted by smooth lens models.       
Even a clump as small as a star can perturb the    
image of small sources ($\sim      
100$ AU), but in this case we would see a microlensing effect such as      
variability in the image brightness with a time scale of order      
months.    
Some authors tried to fit systems (for example the radio system B1422+231) with anomalous flux ratios using
  smooth lens and multipole models: although it seems necessary to
  investigate further whether the multipoles can fit the lens configurations, these methods are not exhaustive (Evans \&
  Witt 2003, {  Kochanek \& Dalal 2004}, Congdon \& Keeton 2005) and it is not yet possible to conclude
  that the multipole
  approach can explain the anomalous flux ratios.
{  Indeed, Kochanek \& Dalal 2004 showed that the flux
 anomaly
distributions display the 
characteristic demagnifications of the brightest saddle point relative
 to the
other images expected for low optical depth substructure, which they conclude cannot
 be
mimicked by problems in the ``macro'' models for the gravitational potential of the lens galaxy.}
Mao \& Schneider (1998), Keeton (2001), Metcalf \& Madau      
(2001), Brada\v c et al. (2002), Dobler \&Keeton (2006)  noted that a simple way of solving the
puzzle was to put a satellite near the images, and they found
that this could explain the anomaly in B1422+231.

Generally the flux ratios between the      
images do not depend on wavelength since       
they are independent of the intrinsic flux and       
variability of the source (Keeton et al. 1997,   
Mao \& Schneider 1998, Keeton 2001, Metcalf \& Zhao 2002).      
Such discrepancies would probably be due to sub-lensing or      
microlensing effects.      
On the other hand when modeling a multiple QSO lens system, one can either include or      
disregard the flux ratios of the images.        
As pointed out by Chang \& Refsdal (1979) ({  and in the following by
  several authors, as for instance Metcalf 2005, Keeton et al. 2005, Mortonson et al. 2005}) we have to pay attention to      
the fact that the projected (on the lens plane) sizes of the optical      
continuum emitting regions of QSOs are expected to be of the same order      
as the Einstein radius of a star in the lens galaxy ($\sim$ 100AU),    
so that the      
optical magnitudes may well be affected by gravitational microlensing      
(see {  Metcalf et al. 2004, Mortonson et al. 2005, and references therein)}, even if averaged over      
long periods of time.       
The radio and mid-IR regions, when projected on the lens plane,   
are typically of the order      
of 10 pc and change in their magnification should be      
dominated by larger scales than stars (Metcalf 2005, Chiba et al. 2005). If    
the lens galaxies contain      
substructures with an Einstein radius comparable or greater than the projected      
size of the radio component (corresponding   
for the substructure to masses $\geq 10^{8} M_{\odot} $) we should see      
image splitting and distortions {  if they lie close enough to the images (see
section 3.2}, Wambsganss \&      
Paczy\'nski 1994, Mao \& Schneider 1998),      
which have not yet been detected (this might be   
the case for B0128+437, Biggs et al. 2004).      
      
The existence of      
anomalous fluxes in many lens systems has been known since some time.       
The substructure lensing effects have been studied      
by considering single substructures    
(Mao \& Schneider 1998, Metcalf \& Madau 2001),    
by assuming a statistically      
distributed sample of substructures (Chiba 2002, Chen et al. 2003, Keeton et  
al. 2005)  or by simulations (Amara et al. 2006, Macci\`o et al.
  2005).\footnote{These works deal with violations of the cusp relation.}     
An explanation by lensing of substructures for the anomalous       
flux ratio for the PG1115+080 system has      
already been considered (Chiba 2002, Dalal \&      
Kochanek 2002, Chen et al. 2003, {  Kochanek \& Dalal 2004}, Keeton, Gaudi,
Petters 2005), however, mainly within a statistical treatment to      
determine whether a plausible collection of mass clumps could      
explain the strange flux ratio.      
As pointed out by Chiba (2002), even if   
it seems difficult to reproduce anomalous     
flux ratios with CDM subhalos, he concluded that   
the main role in reproducing the observed flux ratio is    
played by the one satellite which is located in the vicinity   
of an image (either A1 or A2 in PG1115+80).

In this paper we analyze in detail the lens     
system by adding one or two subclumps nearby one of the images and     
by solving the lens equation.      
In section 2 we review the main observation and analysis done so far    
on the two lens systems PG1115+080 and B1555+375.     
In section 3 we briefly recall the relevant formalism for   
gravitational lensing and how we proceed when we consider a   
lens model with a perturbation   
induced by one or more substructures.   
Assuming a SIS model for the substructures we can then get an estimate   
on their position and Einstein radius (or mass) such as to    
modify the flux of the image pair near the critical curve due to a source   
located close to a fold.   
In section 4 we present the numerical      
simulations and fits to the two considered lens systems.     
We conclude with a short summary and discussion of our results in section 5.   

\label{1} 

\section {About PG1115+080 and B1555+375}      
\label{sec:data}

PG1115+080 is the second gravitationally lensed quasar which was   
discovered (Weymann et al. 1980, Impey et al. 1998 and references therein).       
The source is at redshift $z_s=1.722$ and the lens galaxy at $z_l=0.310$.       
It is an optically selected, radio-quiet quasar.      
Hege et al. (1981) first resolved the four quasar images      
(a close pair A2/A1, B      
and C), confirming the early model of Young et al. (1981) that the      
lens is a five-images system, one image being hidden by the core of      
the lens galaxy.       
        
Young et al. (1981) noted that the lens galaxy seems      
to be part of a small group centered to the southwest      
of the lens, with a velocity dispersion of approximately $270\pm70$ km      
s$^{-1}$ based only on four galaxy redshifts ({  Tonry 1998 also
confirmed the group velocity dispersion by using five galaxies and getting a
value of 326 $kms^{-1}$}).        
The group is an essential component of any model       
to successfully fit       
the lens constraints (Keeton et al. 1997; Schechter et al.      
1997).      
Also two time delays between the images       
were determined by Schechter et al. (1997)      
and confirmed by Barkana (1997).       
Their results were analyzed by Keeton \& Kochanek (1997) and Courbin       
et al. (1997) {  by assuming power law mass distributions} and leading to a value of       
$H_0=53_{-7}^{+15}$ km s$^{-1}$ Mpc$^{-1}$,       
with comparable contributions to the uncertainties both from the time delay      
measurements and the models.
{  Recently, Read et al. 2007, assuming a non parametric mass distribution, found 
  $H_0= 64_{-9}^{+8} $ km s$^{-1}$ Mpc$^{-1}$ consistent with the currently 
accepted value of about 70 km s$^{-1}$ Mpc$^{-1}$.}

We take the data for the PG1115+080 system       
from Impey et al. (1998) (their Fig.1 and Tables 1 and 2)       
who presented a near-infrared observation of the PG1115+080 system      
obtained with the Hubble Space Telescope (HST) NICMOS camera.       
The flux ratio of the close pair of images (A1 and A2, see Fig. 1)   
is approximately $0.67$ and showed little variation with wavelength 
from the multiple wavelength observations by Impey et al. (1998)
\footnote{Pooley et al. (2006), however, report 
that there has been some variation also in the optical.}.     
Simple lens models require instead an A2/A1       
flux ratio close to $1$, because the images are symmetrically arranged       
near a fold caustic.       
There is no smooth lens model that can explain this anomalous flux:       
while each of such models can differ in        
complexity or in parameterization, the observed discrepancy       
in the flux ratio, compared with the expected universal       
relations for a cusp or fold singularity, suggests that it       
is an intrinsic difficulty for smooth lens models,       
not associated with a particular choice of the parameters  
(Yoo et al. 2005).      

 
 Recently, Chiba et al. (2005) analyzed observations of the PG1115+080 system 
 done in the mid-infrared band and found a flux ratio A2/A1 of 0.93 for the 
 close pair, which is virtually consistent with smooth lens models but 
clearly inconsistent with the optical fluxes.  
The observations indicated that the measured mid-infrared flux originate 
from a hot dust torus around a QSO nucleus. Based on the size estimate of the 
dust torus, they placed limits on the mass of the substructure causing the 
optical flux anomaly\footnote{They also 
  considered for this system the microlensing hypothesis and estimated the time 
  variability of the images flux in the mid-infrared band, which turned out 
to be rather long (more than a decade). This estimate is consistent with the 
fact that the optical flux ratio has remained unchanged over the past decade. 
 It is thus clearly not yet possible to assess the microlensing hypothesis.}.  
For a substructure modeled as a SIS  
the subclumps should have a mass of at most $2.2 \times 10^4 M_{\odot}$ inside a radius 
of 100pc to prevent anomalies in the mid-infrared band. 
However, it has to be pointed out that this 
latter result is based on several assumptions and few observations (Minezaki
 et al. 2004), so that the given 
 value may be subject to substantial modifications. Indeed, if the size of the cooler 
dust torus causing the mid-infrared flux is underestimated then the 
above limit gets increased.
 Furthermore, Pooley et al. 2006 analyzed the system using recent X-ray
 observations, which show also a strong anomalous flux ratio. 
They do not exclude the microlensing hypothesis in order to explain 
the anomaly in the X-ray band, nonetheless they conclude that the
optical emission region should be much larger (by a factor $\approx 10-100$) than predicted
by a simple thin accretion disk model. 
{  Within this model the source size should be $R_s \approx
 10^{15} cm$ (e.g. Wambsganss, Schneider, \& Paczy\'nski 1990; Rauch \& Blandford 1991;
Wyithe et al. 2000). Therefore, if it is 10-100 bigger, $R_s \approx
 0.01-0.1$ pc, the effect of stellar
 microlensing could be ruled out (Metcalf 2005).}

An interesting quadruply imaged lens system is B1555+375   
with a maximum separation of only 0.42 arcsec, which was       
discovered some years ago (Marlow et al. 1999).      
It has an anomalous flux ratio in the radio:   
the system was     
observed at 8,4 GHz at VLA and with MERLIN 5 GHz snapshot observations.      
There are only few observations in    
the optical and near-infrared      
band.   
Marlow et al. (1999) considered a model for B1555+375,    
which describes well the positions   
of the images but fails to reproduce accurately the    
flux ratio between the two images near the    
fold critical point (labeled by A and B, see Fig.3). The observed   
ratio is about B/A $\sim$ 0.57.   
This anomaly has also been discussed by {  Keeton
  et al. (2005) and Dobler \& Keeton (2006)}. As the redshifts of lens and source have not yet been
measured we will adopt  the same values as used by Marlow et al. (1999): $z_l=0.5$ and   
$z_s=1.5$.

\begin{table*}     
      
\begin{tabular}[c]{cccc}      
\hline      
Image & H & I & V \\      
 &mag&mag&mag \\      
\hline

A1   & 15.75$\pm$0.02 & 16.12 & 16.90 \\       
A2   & 16.23$\pm$0.03 & 16.51 & 17.35 \\      
B    & 17.68$\pm$0.04 & 18.08 & 18.87 \\      
C    & 17.23$\pm$0.03 & 17.58 & 18.37 \\      
Lens & 16.57$\pm$0.10 & 18.40 & - \\      
      
\hline      
\end{tabular}      
\centering      
\caption{{Photometric data in 3 bands for the four images of     
PG1115+080, from Impey et al. (1998).}\label{simpar}}      
\vspace{0.2truecm}      
\par\noindent      
\end{table*}

\section{Analytical treatment}      
We briefly recall the general expressions for      
the gravitational lensing and refer, e.g.,     
to the book by Schneider et al. (1992)      
(which we will denote afterwards with SEF)     
and the review by Kochanek (2004).     
The lens equation is       
\begin{equation}      
\label{eq-lens}      
\vec{\beta} =\vec{\theta}-\vec\alpha(\vec\theta)~,     
 \end{equation}      
where $\vec\beta(\vec\theta)$ is      
the source position and $\vec\theta$ the image position.      
$\vec\alpha(\vec\theta)$ is the deflection angle, which depends      
on $\kappa(\vec\theta)$ the dimensionless surface mass density or     
convergence in units of the critical surface      
mass density $\Sigma_{\rm crit}$, defined as      
\begin{equation}      
\label{eq-crit}      
\Sigma_{\rm crit} = { c^2  \over {4 \pi G }} { D_S  \over {D_L D_{LS} }},      
\end{equation}      
where $D_S, D_L, D_{LS}$ are the angular diameter distances between observer and source, observer and lens, source and lens, respectively.

\subsection{Lens Mapping }      
\label{subsec:lens}      
      
In the vicinity of an arbitrary point, the lens mapping       
can be described by its Jacobian matrix $\cal A$:      
\begin{equation}      
       {\cal A} = { \partial \vec \beta \over \partial \vec \theta} =       
         \left( \delta_{ij}       
                - {\partial \alpha_i (\vec \theta) \over \partial \theta_j}      
          \right)      
         =       
         \left( \delta_{ij}       
         - {\partial^2 \psi (\vec \theta)       
                \over \partial \theta_i \partial \theta_j} \right).      
\end{equation}      
Here we made use of the fact (see SEF),      
that the deflection angle can be expressed      
as the gradient of an effective two-dimensional scalar      
potential $\psi$: $\vec \alpha=\vec \nabla_{\theta} \psi$,       
which carries information on       
the Newtonian potential of the lens.      
The magnification is defined as the ratio between the      
solid angles of the image and the source (since the surface      
brightness is conserved) and is given by the inverse      
of the determinant of the Jacobian $\cal A$      
\begin{equation}      
        \label{eq-magn-det}      
        \mu = {1 \over \det {\cal A}}.      
\end{equation}      
   
The Laplacian of the effective potential      
$\psi$ is twice the convergence:        
\begin{equation}      
        \label{eq-kappa}      
 \psi_{11} + \psi_{22}  = 2 \kappa = {\rm tr } \    \psi_{ij} .      
\end{equation}      
With the definitions for the components of the external shear $\gamma$:      
\begin{equation}      
        \label{eq-gamma1}      
  \gamma_1 (\vec \theta) = {1 \over 2} (\psi_{11} - \psi_{22}) =       
  \gamma (\vec \theta) \cos [2 \varphi (\vec \theta)]       
\end{equation}      
and       
\begin{equation}      
        \label{eq-gamma2}      
  \gamma_2 (\vec \theta) =  \psi_{12}   =  \psi_{21} =       
  \gamma (\vec \theta) \sin [2 \varphi (\vec \theta)]       
\end{equation}      
(where the angle $\varphi$ gives     
the direction of the shear)      
the Jacobian matrix can be written as      
\begin{equation}      
  {\cal A} =       
         \left(       
        \begin{array}{cc}       
                        1 - \kappa - \gamma_1 & -\gamma_2 \\      
                           -\gamma_2  &  1 - \kappa + \gamma_1 \\      
        \end{array}       
        \right)      
\end{equation}      
\begin{equation}      
        =  ( 1 - \kappa )      
         \left(       
        \begin{array}{cc}       
                        1 & 0 \\      
                        0 & 1 \\       
        \end{array}       
        \right)      
        - \gamma      
         \left(       
        \begin{array}{lr}       
                        \cos 2\varphi  &  \sin 2\varphi  \\      
                        \sin 2\varphi  & -\cos 2\varphi  \\       
        \end{array}       
        \right),     
\end{equation}      
where $\gamma=\sqrt{\gamma_1^2 + \gamma_2^2}$.     
With eq.(8) the magnification can be expressed as a function of the       
convergence $\kappa$ and the shear $\gamma$ at the image point:      
\begin{equation}      
\mu = ( \det {\cal A})^{-1}  =  { 1 \over (1-\kappa)^2 - \gamma^2}.      
\end{equation}      
Locations at which $\det A = 0$  have formally infinite      
magnification are the critical curves in the lens      
plane. The corresponding locations in the source plane are the      
caustics. For spherically symmetric mass distributions      
the critical curves are circles,      
whereas for elliptical lenses or spherically      
symmetric lenses with external shear,       
the caustics can have cusps and folds.      
      
Near a fold the lens equation      
can be reduced to a one-dimensional model and       
a Taylor expansion can be performed (see SEF, Kochanek 2004), for which we get      
      
\begin{equation}      
\beta -\beta_{0}=\frac{\partial\beta}{\partial\theta}(\theta-\theta_{0})+{1\over2}     
\frac{\partial^{2}\beta}{\partial\theta^{2}}(\theta-\theta_{0})^{2},      
\end{equation}      
i.e.      
\begin{equation}      
        \beta = \theta\left( 1-\psi'' \right) - { 1 \over 2 } \psi''' \theta^2      
             \rightarrow  - { 1 \over 2 } \psi''' \theta^2      
\end{equation}      
and inverse magnification      
\begin{equation}      
        \mu^{-1} = \left( 1-\psi'' \right) - \psi''' \theta       
             \rightarrow - \psi'''\theta.      
\end{equation}      
We choose the coordinate system such that there is a critical     
line at $\theta=0$ (i.e. $1-\psi''=0$) and the primes denote derivatives with respect to $\theta$.      
These equations are easily solved and one finds      
that the two images are       
at $\theta_{\pm}=\pm (-2\beta/\psi''')^{1/2}$      
with inverse magnifications $\mu_{\pm}^{-1} = \mp (-2\beta\psi''')^{1/2}$       
that are equal in       
magnitude but with opposite sign.        
Hence, if the assumptions for the Taylor expansion      
hold, the images merging at a fold should have identical fluxes.       
Using gravity to produce anomalous flux ratios requires terms       
in the potential with a length scale comparable to the separation       
of the images to significantly violate the rule that they should have similar fluxes.

\subsection{Perturbing the system}      
\label{subsec:perturbing}      
Let's consider a general lens system configuration for which we know flux      
ratios and image positions and we assume to be     
able to reproduce with a smooth lens      
model, such as a singular isothermal ellipsoid (SIE), the main features     
of the lens, besides the anomalous flux ratio.     
Adding an external potential term in the lens equation and   
correspondingly in the Jacobian    
matrix such as induced by singular isothermal sphere (SIS)     
substructures perturb the system.    
Keeton (2001) analyzing the system B1422+231 (cusp case), could put   
limits on the subclump mass range by considering the different effects they   
would induce on optical and radio bands.     
For the same system Brada\v c et al. (2002) suggests a way to estimate the     
minimum value for the convergence $k$   
in order to get agreement with the observed flux ratios.      
   
Here we constrain the mass and the position    
of a substructure by considering   
its effects on the flux of the images. At each image position   
the perturbed Jacobian matrix can be written as     
\begin{equation}      
  {\cal A} =       
         \left(       
        \begin{array}{cc}       
                        1 - \kappa_{1} - \tilde\gamma_1 & -\tilde\gamma_2 \\      
                         -\tilde\gamma_2 &  1 - \kappa_{1} - \tilde\gamma_1 \\     
         \end{array}\right),       
         \end{equation}      
where $\kappa_{1}= (\kappa+\Delta{\kappa}), \tilde\gamma_1=      
(\gamma_1+\Delta{\gamma_1}), \tilde\gamma_2= (\gamma_2+\Delta{\gamma_2}$)    
and        
$\Delta{\kappa},\Delta{\gamma_{1}}$ and $\Delta{\gamma_{2}}$      
are the perturbations induced by a substructure.

If the substructure is modeled by a SIS, it is possible to express the shear      
components as a function of $\Delta{\kappa}$ (this is true for models that have radial   
symmetry, Kormann et al. (1994)):     
$\Delta{\gamma_{1}}=\Delta{\kappa}cos\theta_{sis}$ and     
$\Delta{\gamma_{2}}=\Delta{\kappa}sin\theta_{sis}$, where      
\begin{equation}     
\Delta{\kappa}=\frac{R_{sis}}{2\sqrt{(x_{sis}-x_{P})^2+(y_{sis}-y_{P})^2}}.      
\end{equation}     
($x_{sis}$, $y_{sis}$)  is the position of the substructure     
and ($x_P$, $y_P$) is the considered image position.    
$R_{sis}$ is the Einstein radius of the substructure, which depends   
on its mass and the distances (or redshifts) to the lens and the source,   
the latter ones being known quantities.    
The $\theta_{sis}$ is given through the relation    
$tg\theta_{sis}= (x_{sis}-x_{P}) / (y_{sis}-y_{P})$      
and it is the angle between the SIS and the    
considered image position ($x_P, y_P$).    
We first consider a model with one additional substructure located at      
the same distance as the lens. 
Its mass and position have to be determined such that the substructure does not      
significantly (within the observational errors) modify the positions of all the images    
as well as the fluxes of the images which lie far from the      
two ones near the fold critical point. These requirements clearly put constraints on the   
mass and position of the substructure.

For the determination of the magnification of an image      
the additional terms due to the substructure depend only on the position  
$(x_{sis}, y_{sis})$ and      
the mass (Einstein radius $R_{sis}$) of the subclump,    
thus we have 3 unknown quantities (see Appendix).       
We consider only three images, thus getting a system with three equations    
for three unknown quantities, and assume that    
the 4th image is far enough such as not to be perturbed by the substructure.    
We then verify a posteriori that the found solution satisfies this latter assumption    
within the measurement errors.    
It turns out indeed to be the case, as the subclump is   
located far from the 4th image, which is chosen as being the most distant   
one from the two near the fold.   
 The system of equations is non-linear, so that  
  the solution is not unique (see Appendix A). However, some solutions have to be discarded as 
    being not physical (imaginary values or a negative Einstein 
    radius). All acceptable solutions are taken as input parameters for the simulations,   
   as will be discussed in the next Section (see also Fig. 1).

Note that the substructure could produce further multiple images of the original one.    
In our case, where we model the substructures as SIS,   
necessary and sufficient condition for multiple images formation is that    
the Einstein radius (of the subclump) $\theta_{Esub}$ has to be greater than half of $\theta_{I}$,    
the distance of the image from the subclump    
($\theta_{Esub} \geq (1/2) \theta_{I}$) (Narayan \& Schneider 1990).       
In Tables 4 and 7 we give the positions of the substructures and of the    
images for the two considered lens systems as obtained from the simulations 
as discussed in the next section. From these data one easily verifies   
that the Einstein radius of the substructures, as given in Tables   
2 and 6, do not satisfy the above condition, thus ruling out   
the formation of further images.   
As noticed by Keeton (2003), for SIS subclumps    
positive-parity images get always brighter,   
whereas negative-parity images get fainter.      
{  In our case for PG1115+080 (B1555+375)  
A1 (A) is the positive-parity image and A2 (B) the negative-parity one.}      
      
\section{Numerical simulations}      
      
In this section we present our simulations and results.     
We use the gravlens code developed by Keeton (2001)\footnote{The software is    
available via the web site:    
http://cfa-www.harvard.edu/castles}, modeling the main galaxy      
acting as lens (in     
both cases) by a SIE and then by adding an external shear term 
  (which we will in the following denote by $SIE_{\gamma}$)
and/or a    
SIS term to take into account the influence of the group in which the galaxy   
is embedded. Moreover, we add       
one or two substructures to take into account the effects on    
small scales.     
For PG1115+080 we use data in H band (see Table 1) taken from Impey et al.    
(1998), and for B1555+375 the data in the 5 GHz radio band (see Table 6)    
from Marlow et al. (1999).    
We allowed a conservative 1     
$\sigma$ error in the relative x- and y- positions of the image     
components (corresponding to an error of at most {  5 mas}),   
and 1 $\sigma$ error on the values of the fluxes   
(corresponding to a variation by 20\%).     
In each model we have different parameters    
and constraints, and the goodness of    
the fit is given by the $\chi^2$ value, {  evaluated on the image plane ($\chi^2_{img}$), and is a    
sum of different contributions: image positions and fluxes, and main galaxy position\footnote{There is an
  alternate way to define the $\chi^2$ that is evaluated in the
{\it source plane\/} ($\chi^2_{src}$) (e.g., Kayser et al. 1990), which is an approximate
version of $\chi^2_{img}$: when using the minimization within
this approximation the formation of additional images is not excluded,
maybe yielding to a not realistic model.
However, the approximation inherent $\chi^2_{src}$, should
properly be used only if a good model is already known, not in an
initial search for a good model (Keeton 2001).}}.      
   
\subsection{PG1115+080}

\begin{figure}
\begin{center}
\includegraphics*[width=10cm]{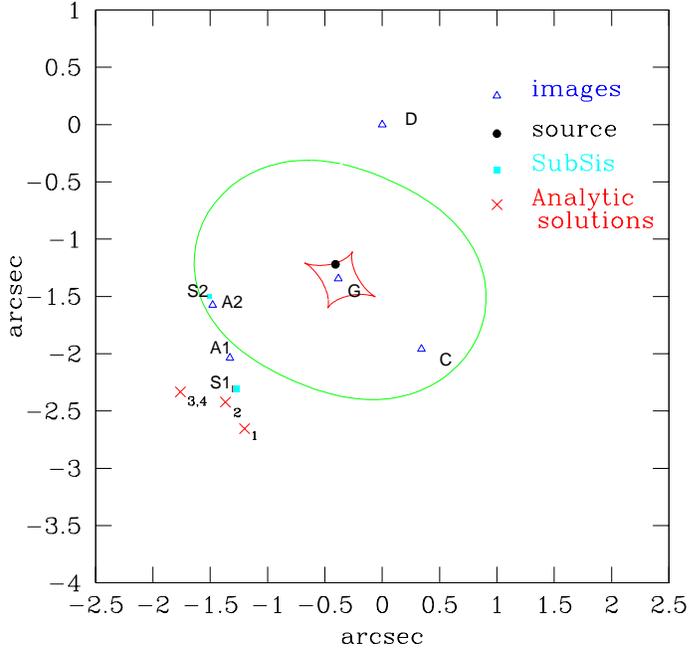}
\end{center}
\caption{PG1115+080: images, source and galaxy (G) 
positions are shown assuming    
a SIE$_\gamma$+SIS model. The position of the     
substructure for both models is also given: note that S1 is closer to the A1
image, while S2 to A2. The critical curve and the caustic   
are for the SIE$_\gamma$ model without the    
modifications induced by the substructure. The analytic solutions as discussed in the Appendix
are also shown. The solutions labeled 
as 3 and 4 are so close that on the figure they
coincide.}
\label{fig:1}
\end{figure}

We model the main lens galaxy as a SIE and take    
into account the contribution      
due to the group of which the galaxy is part of either by adding     
a $SIS_{group}$ component (Keeton 2003, Chen et al. 2003) or an    
external shear term.   We consider both models to which we add
one substructure described as a small SIS.

The results for PG1115+080, modeled as a SIE and an external shear or a SIS,   
are given in Tables 2, 3 and 4 ({  in the following we indicate with $M$ the mass 
  inside the Einstein radius)} and in Fig. 1. 
 In both cases we find good   
agreement with previous results (Impey et al. 1998, Chiba 2002).    
In the case $SIE_{\gamma}$ we find for the    
external shear $\gamma$=0.11 and for its direction an angle of $\phi=56^0$,   
which agree well with the results of Chiba (2002).   
   
In a further step we add one substructure,     
using as starting parameters for its position and mass   
the analytic results as determined following   
the method outlined in the previous section \footnote{Moreover, since we don't 
know   a priori the source flux, we take the value we get from the $SIE_{\gamma}$ model 
and, considering the observational fluxes with respect to the A1 image, we use 
them  as starting values in the analytical system.}.  
In the model $SIE_{\gamma}$ + $SIS$ we have 11 parameters, i.e.      
for the main galaxy : Einstein radius (i.e. mass),      
ellipticity e and orientation PA, shear $\gamma$ and its direction $\phi$;   
for the substructure: the     
Einstein radius, position (corresponding to 2 parameters)   
and, moreover, the source position (2 parameters) and its flux.    
The observational     
constraints are 12, namely the 4 x 2 image positions and 4 fluxes.   

In the model SIE+SIS$_{group}$+SIS, 
in which the group is modeled by a
SIS, the substructure (denoted in Fig.1 by S2) 
is close to A2, whereas for the previous model it is
close to A1 (denoted as S1 in Fig. 1).
  By adding one substructure     
the value of the anomalous flux ratio   
improves substantially for both models: getting lowered from 0.91 to 0.69 for the first model
and to 0.66 for the second model, 
with 
$\chi^2_{tot}$=1.3 and $\approx$ 0.4, respectively, which is quite good. 
\footnote{By further adding a second substructure to the first model,
we find even a lower value for the flux ratio, however,  $\chi^2_{tot}$ and 
$\chi^2_{flux}$ increase, which indicates that this model does not correspond 
to a global minimum, but rather to a local minimum of the $\chi^2_{tot}$ surface.}

\begin{table*}      
      
\begin{tabular}[c]{cccccc}      
\hline       
      
Parameter & SIE$_{\gamma}$ & SIE+SIS$_{group}$&SIE$_\gamma$+SIS&
 SIE+SIS$_{group}$+SIS\\      
 & &   \\      
\hline

  $R_{E,gal}$                 &1.03&  1.14 & 1.12 &1.03\\       
$M_{gal}$($M_{\odot}$) &  $1.23\times 10^{11}$ &$1.05\times 10^{11}$& $1.24 \times 10^{11}$ & $1.05\times 10^{11}$\\      
$R_{E,group}$  & --&2.30    &  --     & 2.11     \\    
$M_{group}$ ($M_{\odot}$) &--  & $4.10\times 10^{11}$  &-- &$4.0\times 10^{11}$\\ 
$R_{E,sub1}$  &-- & --               & 0.033 &0.001\\      
$M_{sub1}$($M_{\odot}$) &    -- &--   &$1.00\times 10^{8}$  &$1.00\times 10^{5}$  \\      
$\sigma_v^{SIE} (km s^{-1})$  &$232.3$ & $245.4$ &$243.5$&$232.3$\\ 
$\sigma_v^{SIS} (km s^{-1})$  &--& --&$39.2$&$6.8$\\
A2/A1               &  0.91 &0.95&0.69  &0.66\\      
$\gamma$            & 0.11 &-- & 0.11 &-- \\       
$\phi$& $56^0$ &--&$56^0$ &--\\      
$\chi^2$ &77 &3.9&1.30  &0.08 \\      
      
\hline      
\end{tabular}      
\centering      
\caption{PG1115+080: Parameters for different models, with and    
without substructure.  
$\gamma$ and $\phi$ are the values of   
the external shear and its direction. {  $R_E$ is expressed in arcsec
  (1 arcsec = 3.19 kpc $h^{-1}$). The 1D velocity dispersions for the main
  component and substructures are also reported}.    
By adding a substructure the agreement between predicted    
and observed fluxes increases substantially.}
  
\vspace{0.2truecm}      
\par\noindent      
\label{table01}
\end{table*}      
   
\begin{table*}      
      
\begin{tabular}[c]{cccccc}      
\hline      
Model&Image &$\kappa$& $\gamma$ & $\mu$& A2/A1  \\      
 & & & & &\\      
\hline

SIE$_{\gamma}$ & A1                  &  0.498& 0.421& 13.35& 0.91 \\       
                  & A2                  & 0.535& 0.545& -12.17&--   \\

SIE+SIS$_{group}$    & A1                  &0.534&0.411&20.21& 0.95\\      
                   & A2                  &0.551&0.504&-19.31&--\\

SIE$_\gamma$+SIS   & A1                 &0.554 & 0.372& 16.54 &0.69\\      
                   &  A2                 &0.561 & 0.531& -11.54&--\\

SIE+SIS$_{group}$+SIS   & A1                 &0.531 & 0.410& 19.46 &0.66\\      
                   &  A2                 &0.565 & 0.517& -12.83&--\\      
\hline      
\end{tabular}      
\centering      
\caption{{ PG1115+080: Values of shear, convergence and amplification     
for A1 and A2 images from simulations for the considered models.}\label{simpar}}      
\vspace{0.2truecm}      
\par\noindent      
\end{table*}

\begin{table*}      
      
\begin{tabular}[c]{ccccc}      
\hline      
Object& x  & y  & e & PA  \\      
&(arcsec)&(arcsec)& \\  
\hline      
Galaxy& -0.381&-1.345 & 0.14& $-84^0$\\      
A1&-1.328&-2.037&--&--\\      
A2&-1.478&-1.576&--&--\\        
Sub1& -1.33 & -2.20&--&--\\     
Sub2& -1.52 &-1.57&--&--\\ 
\hline      
\end{tabular}      
\centering      
\caption{{PG1115+080: positions of the lens galaxy center, the close pair   
A1 and A2, as well as the substructure with respect to the C image (see  
Fig.1). Also ellipticity e and orientation PA of the semi major axis   
with respect to  
x-axis (as measured from East to North and centered in the C image) are given.  
The distances between the substructure and the images A1 and A2   
are bigger than twice their corresponding Einstein radius.  
Thus no further images will be formed.}}      
\vspace{0.2truecm}      
\par\noindent      
\end{table*}

The results from the simulations    
agree, as expected, quite well with the analytical ones.   
Since the mass of the substructures is very small   
as compared to the mass of the lens galaxy, the approximation   
used in the analytical approach to neglect the influence    
induced on the positions of the images   
by the substructures is quite well fulfilled.    
We checked this, indeed, on the results obtained from the   
numerical simulations.   
 
{  Since the positions and magnifications of the images are only
known within a certain accuracy, we computed the corresponding
$1\sigma$ and $2\sigma$ ranges for the value of the mass of the substructure.
Starting from our best model we consider two approaches.
In the first we let all parameters (i.e. the main galaxy ones and the
position of the substructure) vary, whereas in the second one we keep the main galaxy
parameters fixed at the values given by the best fit model and let the
remaining parameters vary. 
The high values for the total $\chi^2$ are due to the bad galaxy position
fit (in
the first case) and to the bad image position fit (in the second case).  
In Fig. 2 we report $\chi^2$ as a
function of the substructure Einstein radius for the
SIE+SIS$_{group}$+SIS model for both cases mentioned above. 
We do not consider larger values for the Einstein radius
as this would correspond to masses for the substructure  of 
order $M(<R_{E})\approx 10^{9}-10^{10} M_{\odot}$, too big to be realistic and 
for which one would see effects on the image position or even
image splitting. 
On the other hand we can also exclude Einstein radius that are too small.
In fact, if it is true that with a standard accretion disk model we get a source
size radius $R_s \approx 10^{15}cm$ (Chiba et al 2005) and that for the
PG1115+080 system the real source size should be about 10-100 times bigger (Pooley et
al 2006), we can estimate roughly the limit of the Einstein radius for which the
effects on the images become negligible.
For a stellar $R_E \approx 10^{-5}$ arcsecs, that corresponds to 0.03 pc
(on the lens plane), there would be no (or little) effect on an image of a source with $R_S \approx 0.1$ pc.  
The minimum value of the curve corresponds to an Einstein radius $\approx
0.001$ arcsec.
Anyway, the curve within the 2$\sigma$ range is rather constrained (in both cases), leading thus 
in practice to a small degeneracy, with a rather narrow range 
for acceptable values of the substructure mass.
The $2\sigma$ range is within an Einstein radius of 0.0005 and $\approx 0.005$ arcsec
corresponding  to
 $\sim 2.5\times 10^4~M_{\odot}$ and $\sim 2.5 \times 10^6~M_{\odot}$.

The $2\sigma$ range for the $SIE_{\gamma} +SIS$ model is somewhat larger.
Considering for example the first case, the Einstein radius for the substructure
could be as high as 0.1 arcsecs, leading to quite a big mass ($\approx 10^9 M_{\odot}$).
}



\begin{figure}
\begin{center}
\includegraphics*[width=7cm]{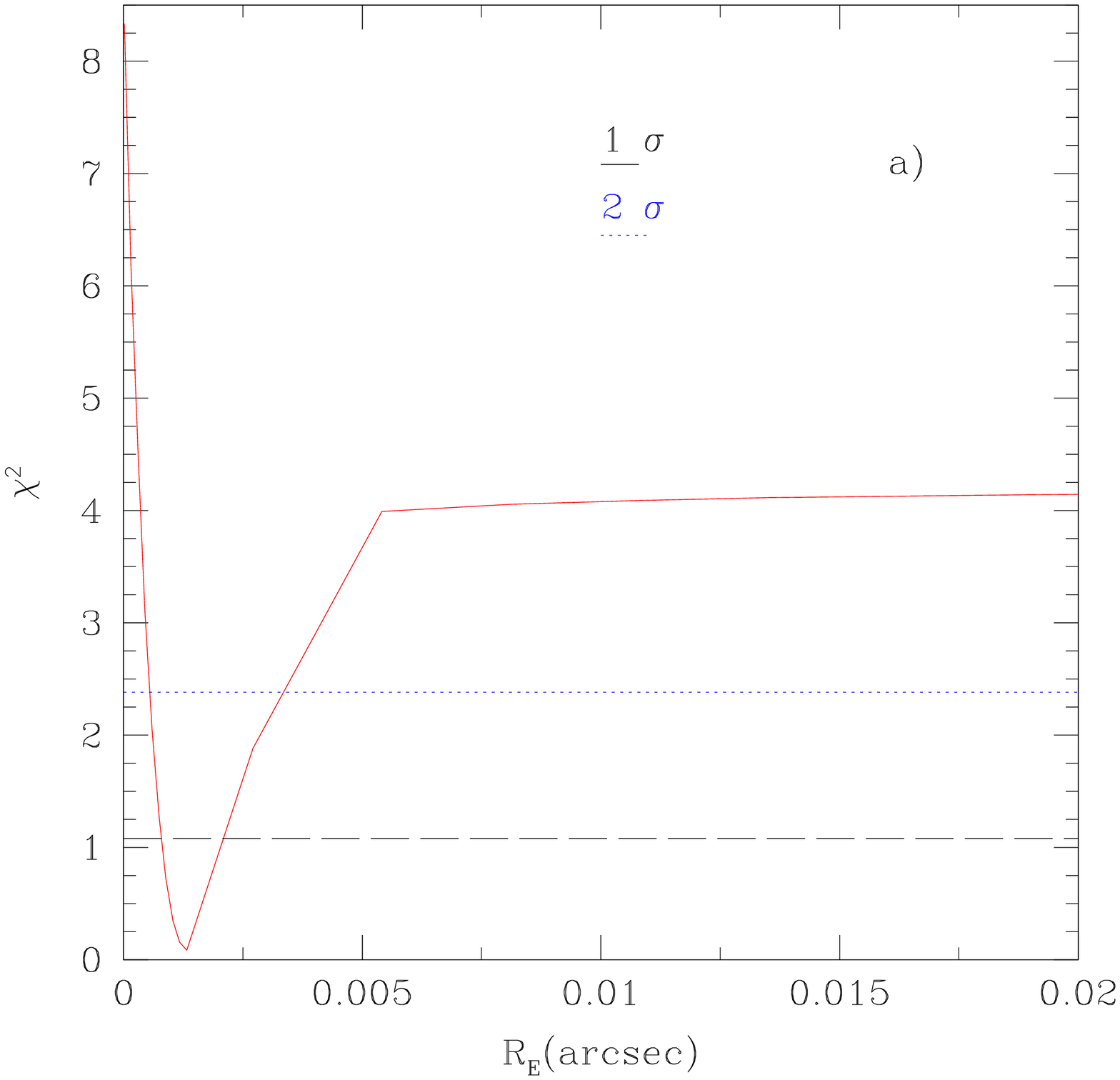}
\includegraphics*[width=7cm]{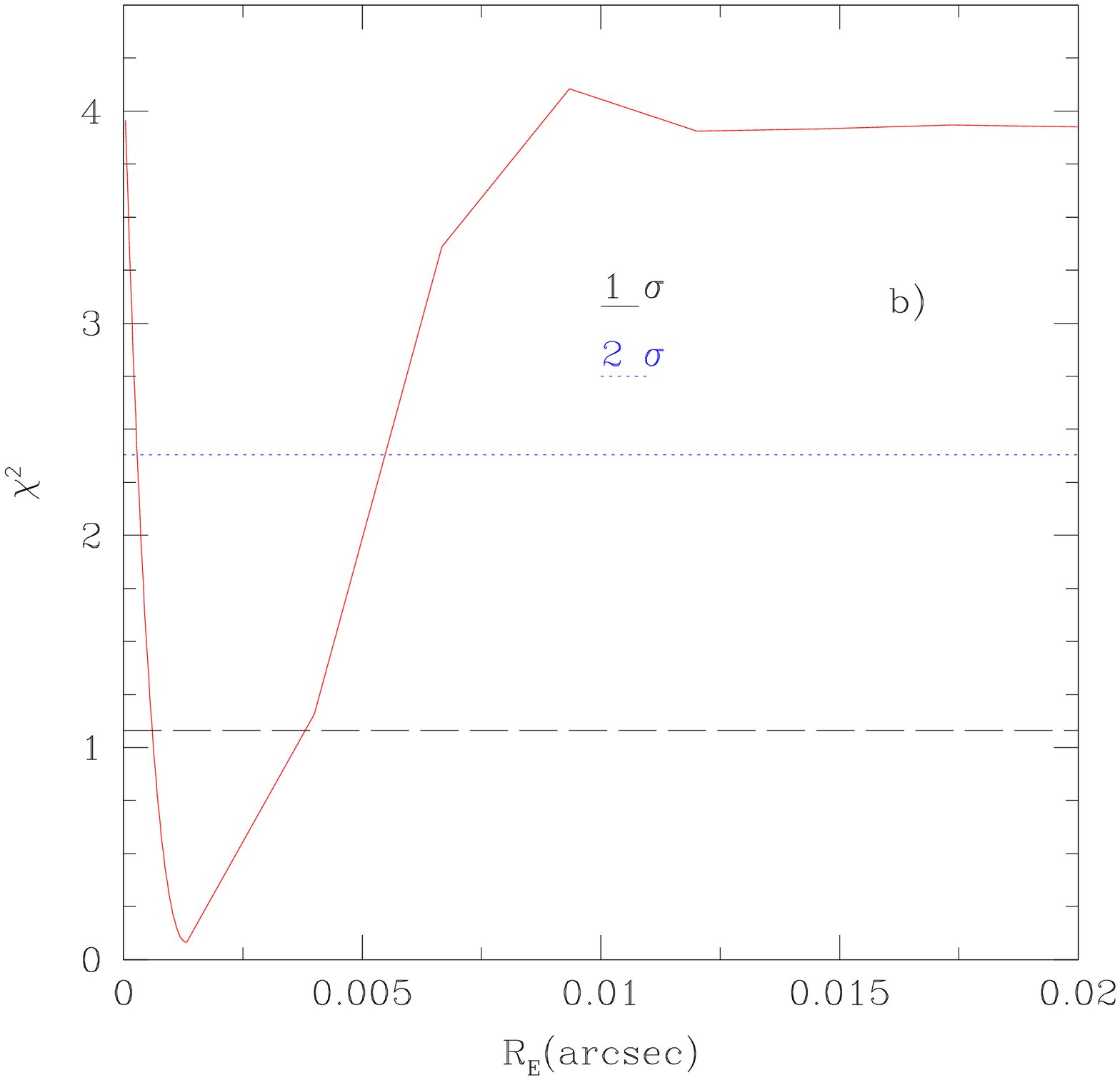}
\end{center}
\caption{1$\sigma$ and 2$\sigma$ limits for the
Einstein radius for the model SIE+SIS$_{group}$+SIS:
a): letting all parameters fixed but the position of the subclump;
b): letting all parameters vary.
}

\label{fig:3}
\end{figure}


\subsection{B1555+375}      
      
  B1555+375 is another lens system for which     
the agreement between observations and model can be improved. 
Already by adding an external shear one can quite 
substantially improve the 
flux fitting, assuming for the lens galaxy a SIE model. Anyway, Dobler 
\& Keeton (2006) consider this model unphysical (since ellipticity and shear 
turn out to be almost perpendicular) and discuss other models,  
adding substructures and giving lower limits on their masses. They find an acceptable model   
using two substructures in front of B and C images, respectively.
 Adding one substructure to the model $SIE_{\gamma}$ does not 
improve the fit further\footnote{We tried, as in the 
  previous case, also 
  analytically to estimate mass and position for a single SIS added to a SIE model, 
but the system of equations does not have in this case real solutions.}.  
On the other hand, adding two substructures (modeled as SIS with comparable masses and located near the pair   
of images due to the fold) to the simple SIE model,    
we get a value for the flux ratio B/A of 0.59, which is only   
slightly higher then the observed one of 0.57 
The results are reported in Tables 6, 7 and 8 and in Fig. 3.      
For the model of a $SIE_{\gamma}$ we get $\chi^2=1.29$ and 
  $\chi^2_{flux}=0.35$. 
{  For the solution with two substructures we find 
a $\chi^2\approx$ 5, with $N_{dof} =5$. 
}      

As we pointed out above, in the radio band    
up to now the only successful explanation for the flux anomalies      
is to consider substructures.    
 {  Moreover, we notice that the SIE model does   
neither fit very well the fluxes of the   
other images besides the close pair ones}. 
 
{  As mentioned in the $SIE_{\gamma}$ model we get a rather high value for 
  the ellipticity (0.85) and for the external shear (0.23): such a strong shear
can be induced by a group of galaxies located 
  around the main lens.
One has also to consider possible effects due to groups of galaxies 
which lie on the line-of-sight, both in the foreground and in the background (Williams et al. 
2006). Other systems (like B1608+656 or HST 12531-2914) 
 show such a high value for the external shear (see Witt \& Mao 1997). In 
  particular, B0128+437 seems to be quite similar to B1555+375 (see Philipps et al. 
1999). Anyway, there are still not enough observations neither about the main galaxy 
nor about its environment, so that it is not possible to choose among the different
solutions. However, our solution with  two substructures involves a SIE lens model with an 
angular structure which is in agreement with previous works (Marlow et al. 1999 and Keeton, 
Gaudi \& Petters 2005).}  
Also for this model we computed the confidence intervals
for the various parameters. 
As an example in Fig. 4 we show the contour ellipses for the confidence
intervals of the Einstein radii of the substructures as obtained fixing the
main galaxy parameters.
  
 \begin{table*}      
      
\begin{tabular}[c]{cccc}      
\hline      
Image & x & y & $S_5$ \\      
 &(arcsec)&(arcsec)& (mJy)\\      
\hline      
      
A   & 0.0$\pm$0.005 &0.0$\pm$0.005 & 17.0 \\       
B   & -0.073$\pm$0.005 & 0.048$\pm$0.005 & 9.7 \\      
C    & -0.411$\pm$0.005 & -0.028$\pm$0.005 & 8.3 \\      
D    & -0.162$\pm$0.005 & -0.368$\pm$0.005 & 1.3 \\      
\hline      
\end{tabular}      
\centering      
\caption{{ B1555+375: Positions and photometric data     
of the 4 images as given by CLASS (from Marlow et al. 1999).}\label{simpar}}      
\vspace{0.2truecm}      
\par\noindent      
\end{table*}

\begin{table*}      
\centering      
\begin{tabular}[c]{cccc}      
\hline      
Parameter & SIE & SIE$_\gamma$ &SIE+2SIS\\       
 & & &   \\      
\hline

  $R_{E,gal}$                 &  0.22 & 0.165 & 0.21\\      
$M_{gal}$ ($M_{\odot} $) &  0.50 & 0.45 &$0.50\times 10^{10}$\\   
    $R_{E,sub1}$  &--&--&0.009\\
$R_{E,sub2}$&--&--&0.012\\
$M_{sub1}$($M_{\odot}$)&--&--&$8.1 \times 10^6$\\
$M_{sub2}$($M_{\odot}$)&--&--&$1.4 \times 10^7$\\

$\sigma_v^{SIE} (km s^{-1})$  &$170.1$ & $147.3$ &$165.6$\\ 
$\sigma_v^{sub1} (km s^{-1})$  &--& --&$23.3$\\
$\sigma_v^{sub2} (km s^{-1})$  &--& --&$24.7$\\

B/A               &  0.93 &0.61  & 0.59\\      
e & 0.53 & 0.85 &0.53\\ 
PA & $-2.42^0$ & $-6.39^0$ &$0.41^0 $       \\ 
$\gamma$            & -- & 0.23&--\\ 
$\phi$&-- &$-78^0$&--\\      
$\chi^2$ &13.9 &1.9 & 5.2  \\   
\hline      

\end{tabular}

\centering      
\caption{{  {B1555+375: Results from the simulations for     
two models without substructures and one model with two substructures. (Notice 
that for the latter model the $\chi^2$ is higher, see text).    
The Einstein radii are expressed in arcsec.     
The system is well fitted already by adding external shear}.}\label{simpar}}      
\vspace{0.2truecm}      
\par\noindent      
\label{table02}
\end{table*}

\begin{figure}
\begin{center}
\includegraphics*[width=10cm]{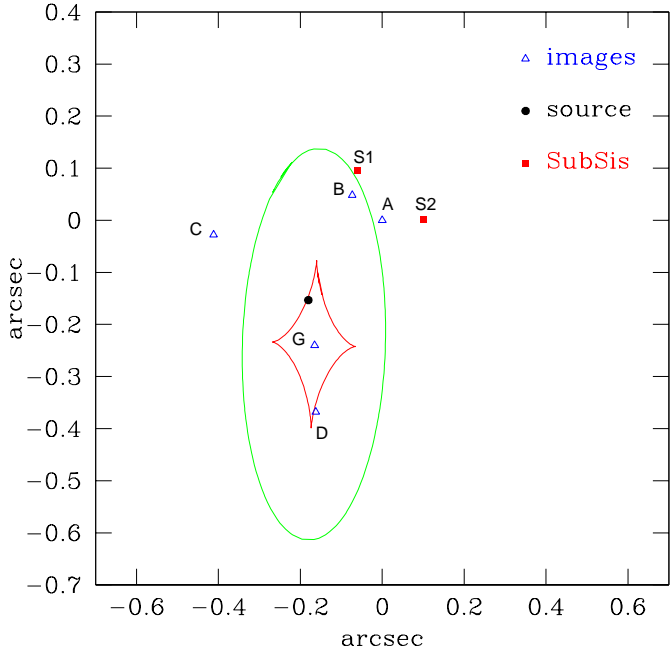}
\end{center}
\caption{B1555+375: images, source and galaxy (G) positions   
are shown assuming   
a SIE+2SIS model. The positions of the two substructures are also   
given. The critical curve and the caustic are for the SIE model alone   
without the modifications induced by the two substructures.}
\label{fig:4}
\end{figure}

\begin{table*}      
      
\begin{tabular}[c]{ccccc}      
\hline      
Object & x & y & e & PA \\   
 &(arcsec)&(arcsec)& \\      
\hline      
      
Lens & -0.162 & -0.246& 0.53& $0.75^0$ \\      
A &0.0 & 0.0 &--&--\\   
B &-0.075 & 0.043 &--&--\\  
   
Sub1  &  -0.060 & 0.094 & --\\      
Sub2  &  0.101 &  0.001 & --\\    
\hline      
\end{tabular}      
\centering      
\caption{{ B1555+375: Parameters of the    
lens model and of the added substructures. e and PA are ellipticity and   
orientation of the semi major axis with   
respect to x-axis (as measured from East to   
North and centered in the A image).}\label{simpar}}      
\vspace{0.2truecm}      
\par\noindent      
\end{table*}

\begin{figure}
\begin{center}
\includegraphics*[width=10cm]{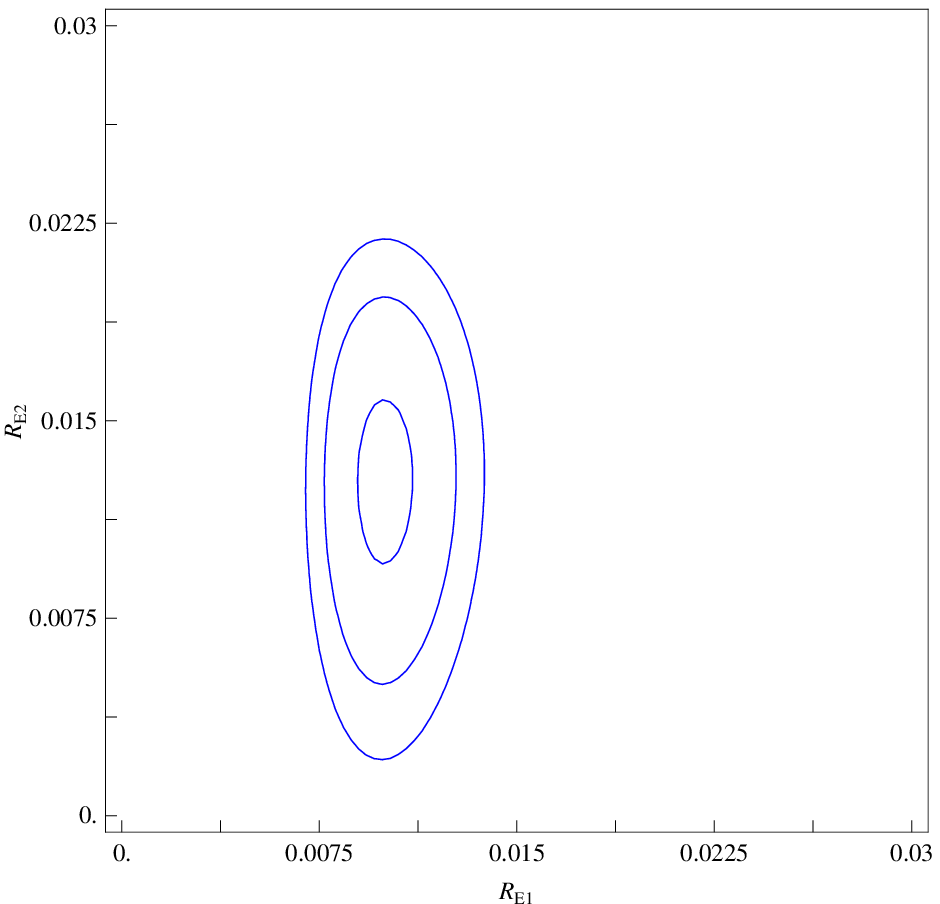}
\end{center}
\caption{  B1555+375:
Contour plot for the $1, 2$ and $3\sigma$ confidence intervals
for the Einstein radii {  (in arcsec)} of the substructures.}
\label{fig:5}
\end{figure}

\section{Discussion}    
   
We have analyzed two lens systems, PG1115+080 and B1555+375,    
which show an anomalous flux ratio for     
the two images near the critical curve, due   
to a fold configuration. These systems cannot   
be modeled using only smooth lens models like SIE,   
although they fit well all the positions of the images.   
We added one or two substructures, taking as starting   
parameters for our numerical simulations the ones obtained   
by an approximated analytical treatment.

In this way we reproduce well, in addition to   
the positions, also the fluxes of the pair   
of images near the critical curve.   
In the PG1115+080 case we get a ratio of 0.69 as compared to the observed    
one of 0.65   if we consider a model SIE$_{\gamma}$ + SIS or 0.66 for a
model SIE+SIS$_{group}$+SIS and for B1555+375 a ratio of 0.59 instead of 0.57 with a
SIE+2SIS model. 
{  We point out that} for PG1115+080 when we consider the model SIE+SIS$_{group}$+
SIS we find that the substructure needed to
  explain the anomaly in the optical band lies close to the A2 image and has a
  mass $\approx 10^5 M_{\odot}$. On the other hand, the
  model SIE$_{\gamma}$ requires a mass $\approx 10^8 M_{\odot}$ close to the
  A1 image that could in
  principle affect also larger $\lambda$ bands,
for which no anomaly has been reported yet (Chiba et al 2005). Therefore, this
latter model is certainly less plausible if not already excluded. 
 Clearly, new observations are still needed to better constrain the models.

{  In the case of B1555+375 we also explored simpler models (i.e. SIE, SIE plus
  simple external shear, or SIE + SIS) but we did not find any
  acceptable solution. 
We note that our best model is similar to the one found by
  Dobler \& Keeton 2006: in their analysis, they concluded that the system B1555+375
  has {\it two anomalous images}.
Their approach is different, since they try to identify anomalies by relaxing
the flux constraints on an image and fit all the others, image positions and
fluxes. When a good model is found, they than conclude that the unconstrained image is
anomalous and give a lower mass limit on the subclump likely responsible for
the anomalous flux by optimizing $\chi^2$ as a function of the subhalo Einstein
radius and the source size. Nevertheless, they also need
  two substructures close to two different images (either in front of images
  A and C or close to images B and C) to fit the fluxes. 
Despite the fact that one of their models has $N_{dof} = 0$,
 the values of the subhalo masses are similar to ours (e.g. $\sim
  10^{5-6} M_{\odot}$ within the Einstein radius).
Finally, we would like to point out that the current observations towards
  B1555+375 seem not to reveal any
nearby group to which the lens system belongs, so a SIE model plus external shear might be inadequate to describe
the effect due to substructures, which are rather better modeled by a SIS.} 

In both cases the best models are the ones with additional substructures.   
Previously, the best models gave values for the anomalous flux ratio   
of 0.91 for PG1115+080 and 0.93 for B1555+375, respectively.   
The improvement achieved by just adding substructures   
is remarkable.   

{  Our approach, although leading to better values for the anomalous flux ratios,
 is not completely new and indeed, as mentioned in the introduction previous works attempted to solve
  the anomalous flux ratios problem by adding perturbations until a good model was found.
In particular, Keeton 2003, Inoue \& Chiba 2005 studied the effect of 
substructures, modeled as SIS, in a convergence and shear field describing the main 
lens galaxy (or galaxies along the line of sight,
e.g. Keeton 2003).
Mao \& Schneider 1998, Metcalf \& Madau 2001, Keeton 2001 analysed the problem
by directly adding subclumps, in form of point masses, SIS, or plane-wave
perturbations, in order to explain the observed anomalies.
Chen 2003, Metcalf 2005, Keeton et al. 2005, Macci\`o \& Miranda 2006 (and
references therein) studied the problem by using statistical approaches and
numerical simulations.}

The masses of the substructures are in the range   
$\approx 10^5 - 10^8~M_{\odot}$, and distant enough from the images not to induce   
the formation of additional ones.    
{  Given the mass range the perturbers could be globular clusters      
or small satellite galaxies, but also CDM dark substructures.  
In order to compare the masses we obtained with dark substructures    
as predicted by CDM, we notice that our values correspond to the mass enclosed
within the Einstein radius, while CDM subhalo masses are defined as the 
mass enclosed within the tidal radius, which is usually much larger
than $R_E$. 
However, without going too much into details, it is still possible to get a rough estimate of the mass
within the tidal radius by calculating the 1D velocity dispersion of the
subhalos from their Einstein radius and then by considering the $M \propto
v_c^3$ (with the approximation $v_c = \sigma_v / \sqrt3$) relation 
found in numerical simulations (Bullock
et al. 2001, Diemand et al 2004). Althought current simulations achieve a lower limit for 
CDM subhalo masses of  $M \approx 10^{10-11}M_{\odot}$, we assume this relation to hold for
lower values of the velocity dispersion and of the masses.
Clearly, the extrapolation to lower velocities could suffer from resolution
effects and numerical noise.
In the system PG115+080 for the $SIE_{\gamma}+SIS$ model the substructure has a $\sigma_{v} \approx 39
km/s$, which leads to a mass of $\sim 7\times 10^8 M_{\odot}$ (the
mass within the Einstein radius, as given in Table \ref{table01}, is instead $1\times 10^8 M_{\odot}$), 
while in the model
$SIE+SIS_{group}+SIS$ the substructure has $\sigma_{v} \approx 6.8
km/s$ corresponding to $\sim 4\times 10^6 M_{\odot}$ (instead of $1\times 10^5 M_{\odot}$). 
In the system B1555+375 the substructures have $\sigma_{v1} \approx 23~km/s$ and 
$\sigma_{v2} \approx 25~ km/s$ respectively, which leads for both a corresponding mass of about  
$10^8 M_{\odot}$ (whereas, as given in Table 6, the mass within the
Einstein radius is $\approx 10^7 M_{\odot}$). We roughly find that the mass within the tidal
radius is about 10 times bigger than the one within the Einstein radius.}

To get a rough estimate of the number of substructures   
expected to lie close to the images we follow the work of   
Diemand et al. (2004). They compute the two  
dimensional radial number density of subhalos   
inside a galaxy virial radius, from which it is then possible   
to get an estimate of the number of   
substructures inside a small area surrounding an image in a lens system.  
The number of subhalos with a mass greater than $m$ inside an area $A$ at a  
distance $r$ from the center of the galaxy is given by  
(Macci\` o \& Miranda 2006)  
\begin {equation}  
N_A(>m,r) = {{ \langle N_{r_v}(>m_0) \rangle  { {m_0} \over {m}}  N(r) A}   
  \over {\pi r_v^2}}~,  
\label{eq:Nr}  
\end{equation}  
where $\langle N_{r_v}(>m_0) \rangle$ is the average number of subhalos  
with $m>m_0$ inside the virial radius $r_v$ of the galaxy  
and $N(r)$ describes the radial  
dependence of the number of substructures.  
The cumulative mass function of subhalos within the virial radius of an halo   
scales as   $\propto m^{-{1}}$.  
  
  As an example for PG1115+080, we consider    
 $m\geq 10^5M_{\odot}$, ($m_0 \approx 10^7$) and a distance of the images from the center   
$r \approx$ 1.5 arcsec ($ N(r)  \approx 2-6 $, $\langle N_{r_v}(>m_0) \rangle
 \approx 166, r_{vir}=268~kpc$,  {  see simulations G0 and G1 in table 1 of
 Diemand et al. 2004 and} Macci\`o \& Miranda 2006).   
Typically, a substructure is expected to lie at a distance from   
an image comparable to the separation between the close pair,  
which is about $\approx$ 0.5 arcsec.  
We, therefore, consider an area corresponding to a small disc with a radius  
about twice the separation distance, thus $ \approx$ 1.0 arcsec (corresponding at $z_l
 =0.31$ to A$\approx \pi(4.5)^2 kpc^2$).  
With these assumptions we expect 10 to 30 substructures in the   
considered area. If instead we require $m\geq   
10^6M_{\odot}$ then we find about 1 to 3 substructures, and 0.1 to 0.3 for $m\geq   
10^7M_{\odot}$ . From these   
considerations we see that 
the expected number of substructures  
within the CDM model seems, in particular, to be lower than required when
 considering the model with a $10^8M_{\odot}$ mass subclump {  or even
 bigger}, whereas for {  models} with a $10^{6}M_{\odot}$ mass subclump it might be 
in better agreement.

{  For B1555+375 the above considerations apply as well: at $z_l
 =0.5$, $1.0$ arcsec$ = 6.114~ kpc$, thus the area around the image pair is A$\approx
 \pi(6.114)^2 kpc^2$. By using the same values as in the previous case 
 the expected number of CDM subhalos bigger then $10^8$ being close to the
 image pair is between 0.02 and $\sim$ 0.1.  
 Clearly, given the rough approximation used, some of
 them depending on extrapolations of numerical simulations of limited
 resolution, it is not possible to draw firm conclusions.}
 
It is obviously also not possible to distinguish   
between different models, moreover it could be that   
some of the substructures are actually located along the line of sight   
rather than being in the surroundings of the lens galaxy.   
To this respect it is interesting to notice that with future ALMA observations   
one could solve this latter problem as pointed out by   
Inoue \& Chiba (2005). They proposed a method to realize a 3D mapping of CDM   
substructures in extragalactic halos, based on astrometric    
shift measurements (at submillimeter wavelengths)    
of perturbed multiple images with respect to   
unperturbed images, with which   
it should be possible to break the degeneracy between the subhalo mass and a   
position along the line of sight to the image.    
   
Also other explanations of the flux anomaly in multiple QSOlens systems   
have been considered in the literature, however   
the best solution,   seems to be the presence of substructures in the halo of the lens galaxy.    
  
  For the second case (B1555+375) there are two acceptable solutions: with and without any 
  substructure.  
Even if a high shear value seems unplausible, we cannot yet 
  rule out this possibility. A recent work (Williams et al. 2006) shows the 
  importance of galaxy groups along the line of sight that can significantly 
  impact the lens model. High-resolution VLA radio observations could 
  help to constrain the lens model further. On the other hand, starting from the 
  simple SIE model, we find a real good fit for the image positions and fluxes, 
  if we add two substructures, still 
keeping acceptable values for the main lens parameters (in agreement with 
  Marlow et al. 1999).
{  We note also that for the B1555+375 system we assumed the redshifts of the
  source and the lens to be known. Even if these values might not be exact
  the uncertainties do not change the result significantly. 
For instance, if we let the redshifts ($z_s = 1.5$ and $z_l = 0.5$) vary by $\pm 0.3$ and use
a source redshift of $z_s=1.8$ with $z_l=0.2$ we get a mass
 for the first substructure of $3.3 \times 10^6 M_{\odot}(<R_E)$. By considering instead
 $z_s=1.2$ and $z_l=0.8$ the mass is 
 $2.7\times 10^7 M_{\odot}(<R_E)$. All other combinations will give a mass 
value within this range, and similarly for the second substructure.}

Finally, we observe  that the PG1115+080 system is radio-quiet, so that the
microlensing hypothesis can not be ruled out (Pooley et
al. 2006). However, given the different observations at the various wavelengths it
might also be possible that both microlensing and millilensing are at work.
A more accurate analysis about the source size could cast some light in constraining 
the substructure size. Since there are discussions
 about it, more high resolution observations are needed to
 definitely rule out the millilensing or microlensing hypothesis. 
On the other hand, for the B1555+375 system {  anomalies are evident in
  the available radio data}, for which
the most plausible explanation are CDM substructures in galactic halos.

\section*{Acknowledgments}     
We thank A. Macci\`o and M. Sereno for useful discussions.      
{  We also thank the referee for useful
comments that improved the paper.} 
Marco Miranda was partially supported by    
the Swiss National Science Foundation.

%
\appendix 
 \section{Analytic estimates for convergence and shear due a substructure} 
 
We briefly present here the formalism used for the analytical approximation
of the total convergence $\kappa_{tot}$ and the total shear 
$\gamma_{tot}$ in the presence of one perturber, located in the main lens plane. 
At each point of the lens plane we can evaluate the total amplification using 
the quantities: 
\begin{eqnarray} 
  \kappa_{tot} = \kappa_{sie}+\Delta{\kappa},\\  
  \gamma_{1tot} = \gamma_{1sie}+\Delta{\gamma_{1}},\\ 
  \gamma_{2tot} = \gamma_{2sie}+\Delta{\gamma_{2}},\\ 
  \mu^{-1} = (1-\kappa_{tot})^2 - \left[(\gamma_{1tot})^2+(\gamma_{2tot})^2\right]. 
\end{eqnarray}  
$\kappa_{sie}$ is the convergence due to the main lens and 
$\Delta{\kappa}$ is due to the perturber and similarly for the shear $\gamma_{sie}$ and 
$\Delta{\gamma}$ (see Sec. 3.2).  
Dealing with a SIE model allows us to write (see Kormann et al. 1994): 
\begin{equation} 
\kappa_{sie} = \frac{R_{sie}}{2\sqrt{(\frac{2q^2}{1+q^2})(x_{sie}-x_P)^2+(y_{sie}-y_P)^2}},  
\end{equation} 
Where $q$ is the axis ratio of the elliptical model used for the galaxy acting as lens. 
The values for $\kappa_{sie}$, $\gamma_{1sie}$ and $ \gamma_{2sie}$ are 
taken, for instance in the PG1115+080 case,
from the $SIE_{\gamma}$ model (which corresponds to the zero order approximation),
computed in A1 and A2 (see Table 3)
as well as the magnification factors 
with respect to the image A1 (and keeping the 
source flux as obtained from $SIE_{\gamma}$ model). In this way the system  
\begin{equation}            
         \left\{       
        \begin{array}{ccc}       
                        \mu^{-1}_{A1} & =(1-\kappa_{totA1})^2 - \left[(\gamma_{1totA1})^2+(\gamma_{2totA1})^2\right]\\ 
                        \mu^{-1}_{A2} & = (1-\kappa_{totA2})^2 - \left[(\gamma_{1totA2})^2+(\gamma_{2totA2})^2\right]\\    
                        \mu^{-1}_{B}  & = (1-\kappa_{totB})^2 - \left[(\gamma_{1totB})^2+(\gamma_{2totB})^2\right] 
\end{array} 
      \right. 
\end{equation}      
has only three unknown quantities, namely $R_{sis}$ and the perturber position given by 
($x_{sis}$, $y_{sis}$).  
Since the system is non-linear, we get different sets of solutions, some of which  
turn out to be unphysical. The allowed solutions are near the close pair (see Table A1).
We marked their positions on Figure 1. We used then these solutions
as input parameters for the numerical simulation. It turns out
that all converge to the same model $SIE_{\gamma}+SIS$ discussed in Sect. 4.1.

\begin{table}     
      
\begin{tabular}[c]{cccc}      
\hline      
&$R_{E}$ & x & y  \\      
&arcsec &arcsec&arcsec \\      
\hline      
      
1&0.030&-1.199 &-2.65 \\
2&0.035&-1.825 &-2.352 \\
3&0.049& -1.363 &-2.421  \\
4&0.050&-1.76&-2.332\\

\hline       
\end{tabular}      
\centering      
\caption{Analytic solutions for PG1115+080 as discussed in the text.}      
\vspace{0.2truecm}      

\par\noindent      
\end{table}


 

\end{document}